\documentclass[a4paper]{jpconf}
\usepackage{graphicx}
\begin{document}
\title{Electromagnetic properties of neutrinos}
\author{Carlo Giunti}
\address{INFN, Section of Turin, Turin, Italy}
\ead{giunti@to.infn.it}
\author{Alexander Studenikin}
\address{Department of Theoretical Physics,
Moscow State University, Moscow, Russia}
\ead{studenik@srd.sinp.msu.ru}

\begin{abstract}
A short review on electromagnetic properties of neutrinos is
presented. In spite of many efforts in the theoretical and
experimental studies of neutrino electromagnetic properties, they still
remain one of the main puzzles related to neutrinos.
\end{abstract}

{ \it Neutrino electromagnetic vertex function.} The neutrino
electromagnetic properties (see \cite{GiuStuPAN09} for a recent
review on this topic and the corresponding references on original
papers) are determined by the neutrino electromagnetic vertex
function $\Lambda_{\mu}(q,l)$ that is related to the matrix element
of the electromagnetic current between the neutrino initial state
$\psi (p)$ and final state $\psi (p')$,
$<{\psi}(p^{\prime})|J_{\mu}^{EM}|\psi(p)>= {\bar
u}(p^{\prime})\Lambda_{\mu}(q,l)u(p)$, where
$q_{\mu}=p^{\prime}_{\mu}-p_{\mu}$,
$l_{\mu}=p^{\prime}_{\mu}+p_{\mu}$.
Lorentz and electromagnetic gauge invariance imply
\cite{GiuStuPAN09} (see also
\cite{KayPRD82,NiePRD82,NowPasRodEJP05_phys0402058}) that the vertex
function can be written in terms of four form factors,
\begin{equation}
\label{vert_func}\Lambda_{\mu}(q)=
f_{Q}(q^{2})\gamma_{\mu}+f_{M}(q^{2})i\sigma_{\mu\nu}q^{\nu}
+f_{E}(q^{2})\sigma_{\mu\nu}q^{\nu}\gamma_{5}+
f_{A}(q^{2})(q^{2}\gamma_{\mu}-q_{\mu}{\not q})\gamma_{5},
\end{equation}
where $f_{Q}(q^{2})$, $f_{M}(q^{2})$, $f_{E}(q^{2})$ and
$f_{A}(q^{2})$ are charge, dipole magnetic and electric, and anapole
neutrino form factors.
The matrix element of the electromagnetic current can be considered
also between neutrino initial $\psi_{i} (p)$ and final $\psi_{j}
(p')$ states with different masses, $p^2=m_i^2$ and $p'^2=m_j^2$. The corresponding
vertex function can be written in the form
\begin{equation}
\label{Lambda} \Lambda_{\mu}(q)=
\Big(f_{Q}(q^{2})_{ij}+f_{A}(q^{2})_{ij}\gamma_{5}\Big)
(q^{2}\gamma_{\mu}-q_{\mu}{\not
q})+f_{M}(q^{2})_{ij}i\sigma_{\mu\nu}q^{\nu}
+f_{E}(q^{2})_{ij}\sigma_{\mu\nu}q^{\nu}\gamma_{5},
\end{equation}
where the form factors are matrices in the space of neutrino mass
eigenstates \cite{MarSanPLB77_LeeShrPRD77_FujShrPRL80_ShrNP82}.

 In the case of Dirac neutrinos, the assumption of
$CP$ invariance combined with the hermiticity of the electromagnetic
current $J_{\mu}^{EM}$ implies that the electric dipole form factor
vanishes. In the case of Majorana neutrinos, regardless of whether
$CP$ invariance is violated or not, the charge, dipole magnetic and
electric form factors vanish \cite{NiePRD82,SchValPRD81}, $f_Q=f_M
=f_E=0$ (the anapole moment can be non-vanishing, see also
\cite{KobOkuProThePhy72}, as well as transition magnetic and electric
moments). Since Dirac and Majorana neutrinos exhibit quite different
electromagnetic properties, the investigation of neutrino
electromagnetic interactions provides a tool for specifying the
neutrino nature.

{\it Neutrino electric form factor.} It is usually believed that the
neutrino electric charge is zero. This is often thought to be
attributed to gauge invariance and anomaly cancellation constraints
imposed in the Standard Model. In the Standard Model  of $SU(2)_L
\times U(1)_Y$ electroweak interactions it is possible to get
\cite{FooLewVolJPG93_BabMohPRD89} a general proof that neutrinos are
electrically neutral which is based on the requirement of electric
charges quantization. The direct calculations of the neutrino charge
in the Standard Model for massless (see, for instance
\cite{BarGasLauNP72_Car_RosBerVid_ZepEPJC00} and references therein)
and massive neutrino \cite{DvoStuPRD04_DvoStuJETP04}  also prove that,
at least at the one-loop level, the neutrino electric charge is
gauge independent and vanishes. However, if the neutrino has a mass,
the statement that a neutrino electric charge is zero is not so
evident as it meets the eye. As a result, neutrinos may become
electrically millicharged particles
\cite{FooLewVolJPG93_BabMohPRD89}.

The most severe experimental constraints on the electric charge of
the neutrino, $q_\nu \leq 10^{-21} e$, are obtained assuming electric
charge conservation in neutron beta decay $n\rightarrow p+e^-+\nu_e$,
from the neutrality of matter (from the measurements of the total
charge $q_{p}+q_{e}$) \cite{MarMorPLB84}. A detailed discussion of
different constraints on the neutrino electric charge can be found in
\cite{Raf_book96_RafPR99}.

Even if the electric charge of a neutrino is vanishing, the electric
form factor $f_Q(q^2)$ can still contain nontrivial information about
neutrino static properties. A neutral particle can be characterized
by a superposition of two charge distributions of opposite signs so
that the particle's form factor $f_Q(q^2)$ can be non zero for
$q^2\neq0$.  The mean charge radius (in fact, it is the charged
radius squared) of an electrically neutral neutrino is given by
$
{<r_{\nu}^2>}=-{6}\frac{df_{Q}(q^2)}{dq^2}{\mid_{ q^2=0}}$,
which is determined by the second term in the expansion of the
neutrino charge form factor
$f_{Q}(q^2)=f_{Q}(0)+q^2\frac{df_{Q}(q^2)}{dq^2}{\mid_{ q^2=0}}$ in
series of powers of $q^2$.

Note that there is a long standing discussion (see \cite{GiuStuPAN09}
for details) in the literature on the possibility to obtain
(calculate) for the neutrino charged radius a gauge independent and
finite quantity. In the corresponding calculations, performed in the
one-loop approximation including additional terms from the $\gamma-Z$
boson mixing and the box diagrams involving $W$ and $Z$ bosons, the
following gauge-invariant result for the neutrino charge radius have
been obtained \cite{BerCabPapVidPRD00_BerPapVidPRL02_NPB04}:
${<r_{\nu_l}^2>}=\frac{G_F}{4\sqrt{2}\pi^2}\Big[3-2\log\big(\frac{m_l^2}
{m^2_W}\big)\Big]$, where $m_W$ and $m_l$ are the $W$ boson and
lepton masses ($l=e,\mu,\tau$)\footnote{This result, however, revived
the discussion \cite{FujShrPRD04_PapBerBisVidEPJ04_NPB05} on the
definition of the neutrino charge radius.}. Numerically, for the
electron neutrino electroweak radius it yields ${<r_{\nu_e}^2>}=4
\times 10^{-33} \, {cm}^2$. This theoretical result differs at most
by one order of magnitude from the available experimental bounds on
$<r_{\nu_i}^2>$. Therefore, one may expect that the experimental
accuracy will soon reach the value needed to probe the neutrino
effective charge radius.

{\it Neutrino magnetic and electric moments.} The neutrino dipole
magnetic moment (along with the electric dipole moment) is the most
well studied among neutrino electromagnetic properties. A Dirac
neutrino may have non-zero diagonal electric moments in models where
$CP$ invariance is violated. For a Majorana neutrino the diagonal
magnetic and electric moments are zero.

The explicit evaluation of the one-loop contributions to the Dirac
neutrino dipole moments in the leading approximation over the small
parameters $b_i={m_{i}^{2}}/{m_{W}^{2}}$ ($m_i$ are the neutrino
masses, $i=1,2,3$), that however exactly accounts for $a_l=
{m_{l}^{2}}/{m_{W}^{2}}$, leads to the following result
\cite{PalWolPRD82}:
\begin{equation}\label{m_e_mom_i_j}
\begin{array}{c}\mu^{D}_{ij} \\  \epsilon^{D}_{ij}\end{array} \Bigg \}  =\frac{e G_F m_{i}}{8\sqrt {2} \pi ^2}
  \Big(1 \pm \frac{m_j}{m_i}\Big)\sum_{l}f(a_l)U_{lj}U^{\ast}_{li},
\ \ f(a_l)=\frac{3}{4(1-a_l)^3}\Big(2-7a_l+6a^2_l-a_l^3-2a_l^2\ln a_l
\Big),
\end{equation}
where $U_{li}$ is the neutrino mixing matrix.
From (\ref{m_e_mom_i_j}) in the limit $a_l\ll 1$,
the diagonal magnetic moment of a Dirac neutrino is given by
\cite{MarSanPLB77_LeeShrPRD77_FujShrPRL80_ShrNP82}
$\mu^{D}_{ii}=\frac{3e G_F m_{i}}{8\sqrt {2} \pi ^2}
  \approx 3.2\times
10^{-19}
  \Big(\frac{m_i}{1 \ eV}\Big) \mu_{B}$.
On the other hand,
the magnetic moment of a hypothetical heavy neutrino $(m_\ell\ll m_W\ll
m_\nu )$ is given by
\cite{DvoStuPRD04_DvoStuJETP04}
$\mu_{\nu}={\frac{eG_{F}m_{\nu}}{8\sqrt{2}\pi^{2}}}$. Note that much
larger values for the neutrino magnetic moments can be obtained in
various extensions of the Standard Model (see, for instance,
\cite{GiuStuPAN09}).

{\it Bounds on neutrino magnetic moments.} Constraints on the
neutrino magnetic moment have been obtained in laboratory $\nu-e$ scattering
experiments from the observed lack of distortions of the
recoil electron energy spectra. Upper bounds on the neutrino
magnetic moment have been obtained in recent reactor
experiments: $\mu_\nu \leq 9.0 \times 10^{-11}\mu_{B}$ (MUNU
collaboration \cite{Dar_eaPLB05}), $\mu_\nu \leq 7.4 \times
10^{-11}\mu_{B}$ (TEXONO collaboration \cite{Wong_PRD07_75_012001}.
The best world limit $\mu_\nu \leq 3.2 \times 10^{-11}\mu_{B}$ has
been recently obtained by the GEMMA collaboration
\cite{BedSta13LomCon}. A stringent limit has also been obtained in the Borexino solar
neutrino scattering experiments: $\mu_\nu \leq 5.4 \times
10^{-11}\mu_{B}$ \cite{BOREXINO_08}. Note
that the magnetic and electric transition moments can contribute
to the effective value of $\mu_\nu$
(see Section~3.6 of \cite{GiuStuPAN09}).

{\it Neutrino electromagnetic interaction effects.} If a neutrino has
the non-trivial electromagnetic properties discussed above, a direct
neutrino coupling to photons is possible and several processes important
for applications exist \cite{Raf_book96_RafPR99}. A set of
most important neutrino electromagnetic processes is: 1) neutrino
radiative decay $\nu_{1}\rightarrow \nu_{2} +\gamma$, neutrino
Cherenkov radiation in an external environment (plasma and/or
electromagnetic fields), spin light of neutrino, $SL\nu$ , in the
presence of a medium \cite{SLnu}; \ 2) photon (plasmon) decay to a
neutrino-antineutrino pair in plasma $\gamma \rightarrow \nu {\bar
\nu }$; \ 3) neutrino scattering off electrons (or nuclei); \ 4)
neutrino spin (spin-flavor) precession in a magnetic field (see
\cite{Okun:1986na}) and resonant neutrino spin-flavour
oscillations in matter \cite{LimMarPRD88_AkhPLB88}. The
tightest astrophysical bound on neutrino magnetic moments, $\big(
\sum _{i,j}\mid \mu_{ij}\mid ^2\big) ^{1/2}\leq 3 \times 10^{-12} \mu
_B$, applicable to both Dirac and Majorana neutrinos,
has been obtained from the observed lack of anomalous
stellar cooling due to plasmon decay \cite{Raf_book96_RafPR99}.

{\it Acknowledgements.} One of the authors (A.S.) is thankful to
Eugenio Coccia, Nicolao Fornengo and Luciano Pandola for the
invitation to attend the TAUP-2009 and to all the conference
organizers for their kind hospitality in Rome.


\medskip
\begin{thebibliography}{99}
\bibitem{GiuStuPAN09}
    Giunti C and Studenikin A 2009 {\it Phys.Atom.Nucl.} {\bf 72} {2151},
    arXiv: 0812.3646
\bibitem{KayPRD82} Kayser B
1982 {\it Phys.Rev.} D {\bf 26} 1662
\bibitem{NiePRD82} Nieves J F 1982 {\it
Phys.Rev.} D {\bf 26} 3152
 \bibitem{NowPasRodEJP05_phys0402058} Nowakowski M, Paschos E and
Rodriguez J 2005 {\it Eur.J.Phys.} {\bf 26} 545

\bibitem{MarSanPLB77_LeeShrPRD77_FujShrPRL80_ShrNP82}
Marciano W J and Sanda A I 1977 {\it Phys.Lett.} B {\bf 67} 303
\nonum Lee B W and Shrock R E 1977 {\it Phys.Rev.} D {\bf 16} 1444
\nonum Fujikawa K and Shrock R E 1980 {\it Phys. Rev. Lett.} {\bf 45}
963
\bibitem{SchValPRD81}Schechter J and
Valle J W F 1981 {\it Phys.Rev.} D {\bf 24} 1883
\bibitem{KobOkuProThePhy72} Kobzarev I and Okun L  1972 {\it
Problems of Theoretical Physics} (Moscow: Nauka) p 219
\bibitem{FooLewVolJPG93_BabMohPRD89} Foot R, Lew H and
Volkas R R 1993 {\it J. Phys.} G {\bf 19} 361; {\it ibid.} 1067
[Erratum]
 \nonum
Babu K S and Mohapatra R N 1989 {\it Phys.Rev.} D {\bf 63} 938
\bibitem{BarGasLauNP72_Car_RosBerVid_ZepEPJC00} Bardeen W, Gastmans R and Lautrup B
 1972 {\it Nucl.Phys.} B {\bf 46} 319
 \nonum Cabral-Rosetti L, Bernabeu J, Vidal J and Zepeda A 2000 {\it Eur.Phys.J.} C {\bf 12} 633
\bibitem{DvoStuPRD04_DvoStuJETP04} Dvornikov M and Studenikin A 2004
{\it Phys.Rev.} D {\bf 69} 073001
\bibitem{MarMorPLB84}Marinelli M and Morpurgo G 1984 {\it Phys.Lett.} B {\bf 137} 439
\bibitem{Raf_book96_RafPR99} Raffelt G  1996 {\it Stars as Laboratories
for Fundamental Physics} (Univ. of Chicago Press)
\bibitem{BerCabPapVidPRD00_BerPapVidPRL02_NPB04}
Bernabeu J, Papavassiliou J and Vidal J 2004 {\it Nucl.Phys.} B {\bf
680} 450
\bibitem{FujShrPRD04_PapBerBisVidEPJ04_NPB05}
Fujikawa K and Shrock R 2004 {\it Phys.Rev.} D {\bf 69} 013007 \nonum
Bernabeu J, Papavassiliou J and Binosi D 2005 Nucl. Phys. B {\bf 716}
352
\bibitem{PalWolPRD82} Pal P and Wolfenstein L 1982 {\it Phys.Rev.} D {\bf 25}
766
\bibitem{Dar_eaPLB05} MUNU Collab. (Darakchieva Z {\it et al.}) 2005 {\it Phys.Lett.}
 B {\bf 615} 153
\bibitem{Wong_PRD07_75_012001} TEXONO Collab. (Wong H T  {\it et al.})
2007 {\it Phys.Rev.} D {\bf 75} 012001
\bibitem{BedSta13LomCon} Beda A G {\it et al.} 2009  {\it Particle Physics on the Eve of
LHC} ed. by Studenikin A (World Scientific: Singapore) p 112,
arXiv:09.06.1926.
\bibitem{BOREXINO_08} Borexino Collab. (Arpesella C {\it et al.}) 2008 {\it
Phys.Rev.Lett.} {\bf 101} 091302
\bibitem{SLnu} Lobanov A and Studenikin A 2003 {\it Phys.Lett.} B {\bf 564}
27; {\it ibid.} 2004 {\bf 601} 171 \nonum Grigoriev A, Studenikin A
and Ternov A 2005 {\it Phys.Lett.} B {\bf 622} 199
 \nonum Studenikin A 2008 {\it J.Phys.A:
Math.Theor.} {\bf 41}  164047
\bibitem{Okun:1986na}
Okun L, Voloshin M  and Vysotsky M 1986 {\it Sov.Phys.JETP} {\bf 64},
446
\bibitem{LimMarPRD88_AkhPLB88} Lim C and Marciano W 1988 {\it
Phys.Rev.} D {\bf 37} 1368 \nonum Akhmedov E 1988 {\it Phys.Lett.} B
{\bf 213} 64
\end{thebibliography}
\end{document}